\documentclass[%
 aip,
 amsmath,amssymb,
 reprint,%
]{revtex4-1}

\usepackage{graphicx}% Include figure files
\usepackage{dcolumn}% Align table columns on decimal point
\usepackage{bm}% bold math
\usepackage[utf8]{inputenc}
\usepackage[T1]{fontenc}
\usepackage{mathptmx}
\usepackage{etoolbox}
\usepackage{color}

\usepackage{siunitx}
\usepackage{multirow}
\makeatletter
\def\@email#1#2{%
 \endgroup
 \patchcmd{\titleblock@produce}
  {\frontmatter@RRAPformat}
  {\frontmatter@RRAPformat{\produce@RRAP{*#1\href{mailto:#2}{#2}}}\frontmatter@RRAPformat}
  {}{}
}%
\makeatother
\begin{document}

\preprint{AIP/123-QED}

\title[]
{Detection of chirality-induced spin polarization over millimeters in polycrystalline bulk samples of 
chiral disilicides $\mathrm{NbSi_{2}}$ and $\mathrm{TaSi_{2}}$}
%Sample Title:\\with Forced Linebreak
% Force line breaks with \\
\author{Hiroaki Shishido}
% \homepage{shishido@pe.osakafu-u.ac.jp.}
 \email[Authors to whom correspondence should be addressed: H. Shishido, ]{shishido@pe.osakafu-u.ac.jp.}
% \altaffiliation[Also at]{Department of Physics and Electronics, Osaka Prefecture University, Sakai, Osaka 599-8531, Japan}%Lines break automatically or can be forced with \\
\affiliation{Department of Physics and Electronics, Osaka Prefecture University, Sakai, Osaka 599-8531, Japan%\\This line break forced with \textbackslash\textbackslash
}%
\author{Rei Sakai}%
\affiliation{Department of Physics and Electronics, Osaka Prefecture University, Sakai, Osaka 599-8531, Japan%\\This line break forced with \textbackslash\textbackslash
}%

\author{Yuta Hosaka}
% \homepage{http://www.Second.institution.edu/~Charlie.Author.}
\affiliation{Department of Physics and Electronics, Osaka Prefecture University, Sakai, Osaka 599-8531, Japan%\\This line break forced% with \\
}%

\author{Yoshihiko Togawa}
% \homepage{y-togawa@pe.osakafu-u.ac.jp.}
% \homepage{http://www.Second.institution.edu/~Charlie.Author.}
 \email[Authors to whom correspondence should be addressed: Y. Togawa, ]{y-togawa@pe.osakafu-u.ac.jp.}
\affiliation{Department of Physics and Electronics, Osaka Prefecture University, Sakai, Osaka 599-8531, Japan%\\This line break forced% with \\
}%


\begin{abstract}
We report that spin polarization occurs over millimeters in polycrystalline bulk samples of chiral disilicide $\mathrm{NbSi_{2}}$ and $\mathrm{TaSi_{2}}$. 
As previously demonstrated in the experiments using single crystals of $\mathrm{NbSi_{2}}$ and $\mathrm{TaSi_{2}}$, 
electrical transport measurements allow detection of direct and inverse signals associated with the chirality-induced spin polarization even in the chiral polycrystals. 
The spin polarization signals also appear in nonlocal measurements, in which charge current flows only in the area millimeters away from the detection electrode. 
These data mean that the spin polarization phenomena occur regardless of the presence of crystalline grains in the polycrystals, indicating a robustness and resilience of the chirality-induced spin polarization.
On the basis of the experimental data, we found that the sum rule holds for the spin transport signals. 
A distribution of handedness over the samples was determined on average in the polycrystals.
While the mechanism of preserving the spin polarization over millimeters remains to be clarified, the present study may open up prospects of spin control and manipulation over macroscopic length scales using chiral materials.

\end{abstract}

\maketitle

Chirality is one of the essential concept of the symmetry widely available among chemistry, biology, and physics. It plays an important role in inducing a wide range of chirality-driven phenomena due to the interplay of structural and dynamical chirality that chiral materials exhibit. Recently, a new class of the phenomena called chirality-induced spin selectivity (CISS) has been found in chiral molecules~\cite{Goh11, Xie11}. 
In the CISS, a spin polarization is generated in electrons that go through a chiral molecule. 
The direction of the spin polarization depends on the handedness of the chiral molecules. 
While the CISS mechanism is an open question~\cite{Naa12, Naa15, Wal21}
, the CISS effect has been observed in many kinds of chiral molecules, such as 
DNA~\cite{Ros13, Kum13}, 
peptides~\cite{Dor13, Ket15, Ara17, Kop17, Tas18, Var18} and  
amino acids~\cite{Zha19, Ziv19, Mis19}, 
helicenes~\cite{Kir16, Ket18}, 
and 
chiral molecule motors~\cite{Sud19, Kul20}.

Very recently, the CISS effect was found in chiral inorganic crystals~\cite{Inu20}. The spin polarization occurs in the electrons traveling along the $c$-axis of the monoaxial $\mathrm{CrNb_{3}S_{6}}$ single crystal. Moreover, the spin-polarized state propagates over one micrometer or longer toward a region in which no net electrical charge current exists in the nonlocal setup. 
The reason why such a nonlocal spin polarization survives on macroscopic length scales is a crucial issue not only for clarifying the CISS mechanism but also for developing solid state device applications using the CISS. 
The CISS properties unique to chiral crystals may provide an advantageous and intriguing platform for designing the CISS applications~\cite{Nab20}.

In this connection, the CISS experiments with polycrystalline samples of chiral materials would be very interesting and may shed light on another feature of the CISS. %and challenging. 
We come up with this idea, inspired by the optical characteristics that a natural optical activity (NOA), which is one of the representatives of chirality-induced phenomena, occurs in a solution with chiral substances. In this case, the orientation of chiral materials is not relevant to the occurrence of the NOA. However, it is not clear whether or not the same principle would hold for the CISS since it has been examined so far mainly in highly-oriented chiral molecules and chiral inorganic single crystals. The availability of the CISS in chiral polycrystals would provide distinct advantages in interpreting fundamental aspects of the CISS as well as in fabricating the CISS devices.

%%%%%%%%%%%%%%%%%%%%%%%%%%%%%%%%%%%%%%%%%%%%%%%%%%%%%%%%%%%%%%%%%%%%%%%%%%%%%%%%%%%%%%%%%%%%%%%%%%%%%%
\begin{figure}[b]
\includegraphics[width=1.0\linewidth]{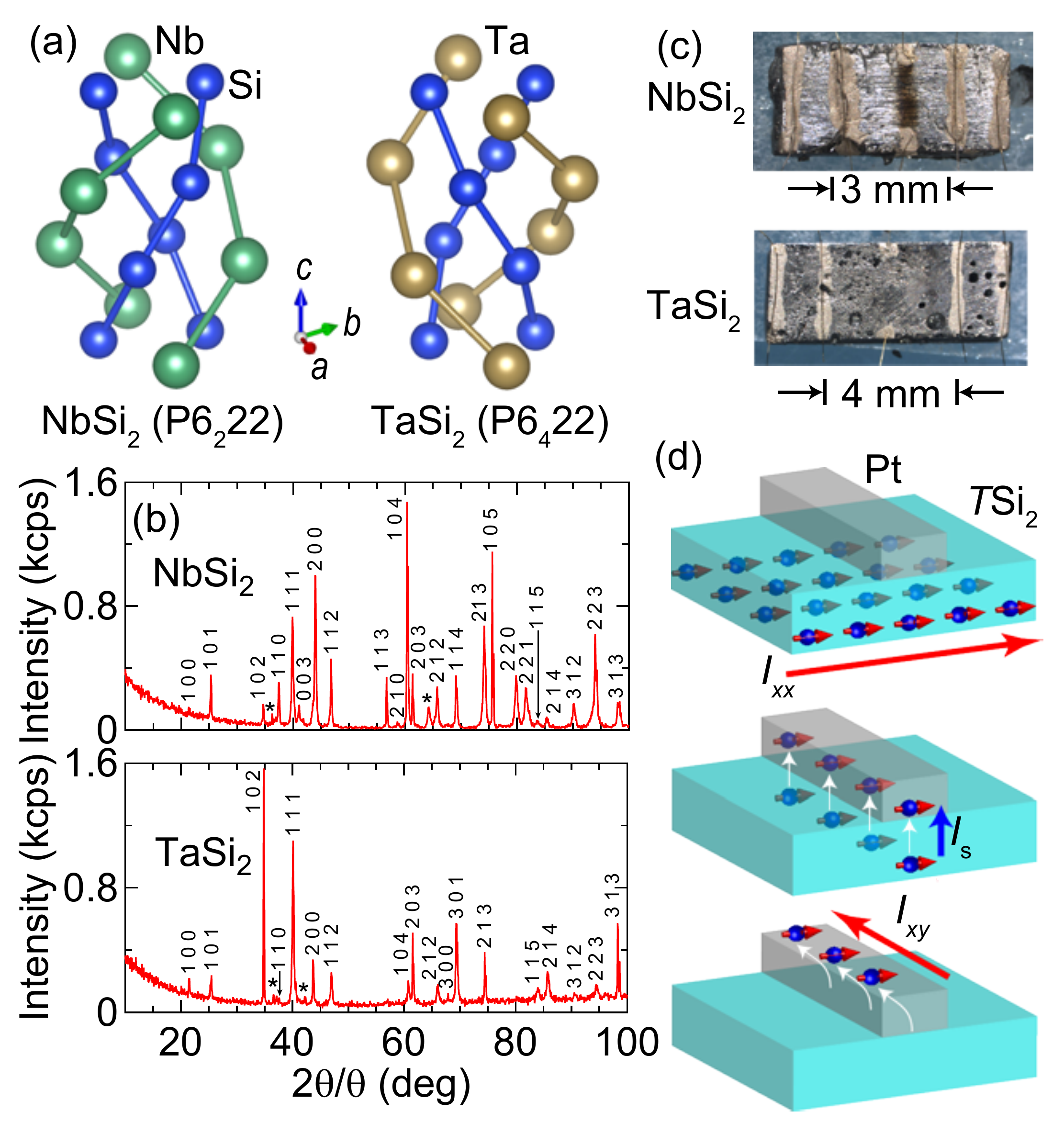}% 
\caption{Schematics of the CISS measurements using chiral disilicide polycrystals of $\mathrm{NbSi_{2}}$ and $\mathrm{TaSi_{2}}$. 
Crystal structures (a), X-ray diffraction patterns of bulk samples (b), photographs of the samples made for the CISS measurements (c) are presented. The procedure of the CISS measurement using a transition metal disilicide TSi$_2$ and platinum (Pt) electrode is schematically drawn in (d). 
}
\label{fig1}
\end{figure}
%%%%%%%%%%%%%%%%%%%%%%%%%%%%%%%%%%%%%%%%%%%%%%%%%%%%%%%%%%%%%%%%%%%%%%%%%%%%%%%%%%%%%%%%%%%%%%%%%%%%%%

In this Letter, we demonstrate the CISS response in polycrystalline samples of chiral disilicide $\mathrm{NbSi_{2}}$ and $\mathrm{TaSi_{2}}$. 
Namely, the spin polarization occurs in chiral bulk polycrystals. 
In a separate experiment, we confirmed that
$\mathrm{NbSi_{2}}$ and $\mathrm{TaSi_{2}}$ single crystals exhibit the CISS response when fabricated into micrometer-sized samples~\cite{Shi21}.
The lengthscale of the present samples, which are made of polycrystals with dimensions of several millimeters in length and width, is hundred times larger than the lamellae of single crystals used in the previous study~\cite{Shi21}. 
Nevertheless, the spin polarization occurs and spreads out over millimeters irrespective of the presence of crystalline grains in the polycrystals.
The present finding indicates a very robust nature of the spin polarization in chiral materials and requires a nontrivial mechanism of protecting the spin polarization on macroscopic length scales.

%%%%%%%%%%%%%%%%%%%%%%%%%%%%%%%%%%%%%%%%%%%%%%%%%%%%%%%%%%%%%%%%%%%%%%%%%
\begin{figure}[t]
\includegraphics[width=1.0\linewidth]{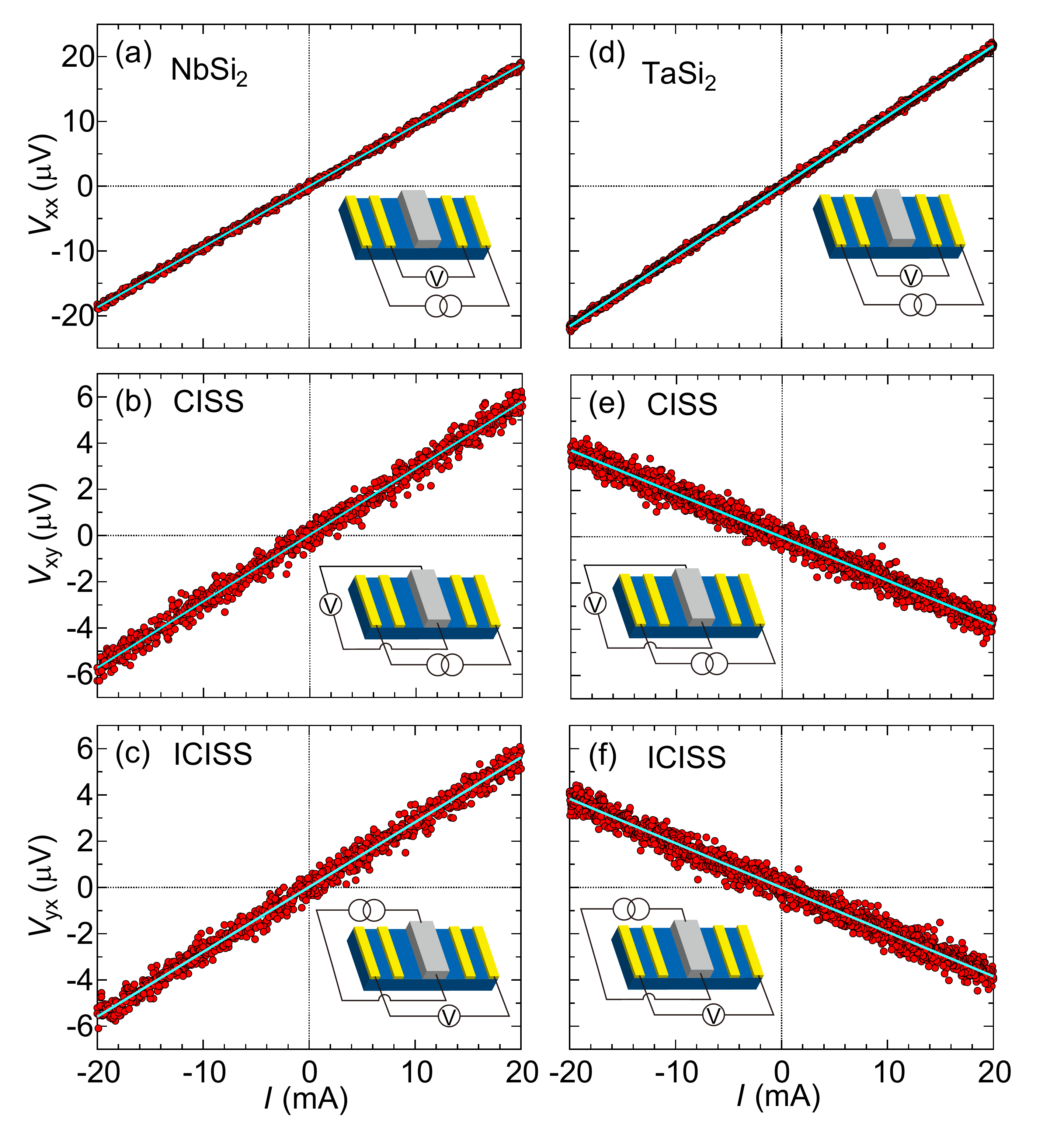}

\caption{
A dataset of the $I$--$V$ characteristics for $\mathrm{NbSi_{2}}$ and $\mathrm{TaSi_{2}}$ polycrystals at room temperature without magnetic fields.
The longitudinal voltage $V_{xx}$, the direct CISS voltage $V_{xy}$ and the inverse CISS voltage $V_{xy}$ are shown in the panels. 
The measurement setup is schematically drawn in the insets. 
}
\label{fig2}
\end{figure}
%%%%%%%%%%%%%%%%%%%%%%%%%%%%%%%%%%%%%%%%%%%%%%%%%%%%%%%%%%%%%%%%%%%%%%%

%%%%%%%%%%%%%%%%%%%%%%%%%%%%%%%%%%%%%%%%%%%%%%%%%%%%%%%%%%%%%%%%%%%%%%%%
\begin{figure}[t]
\includegraphics[width=1.0\linewidth]{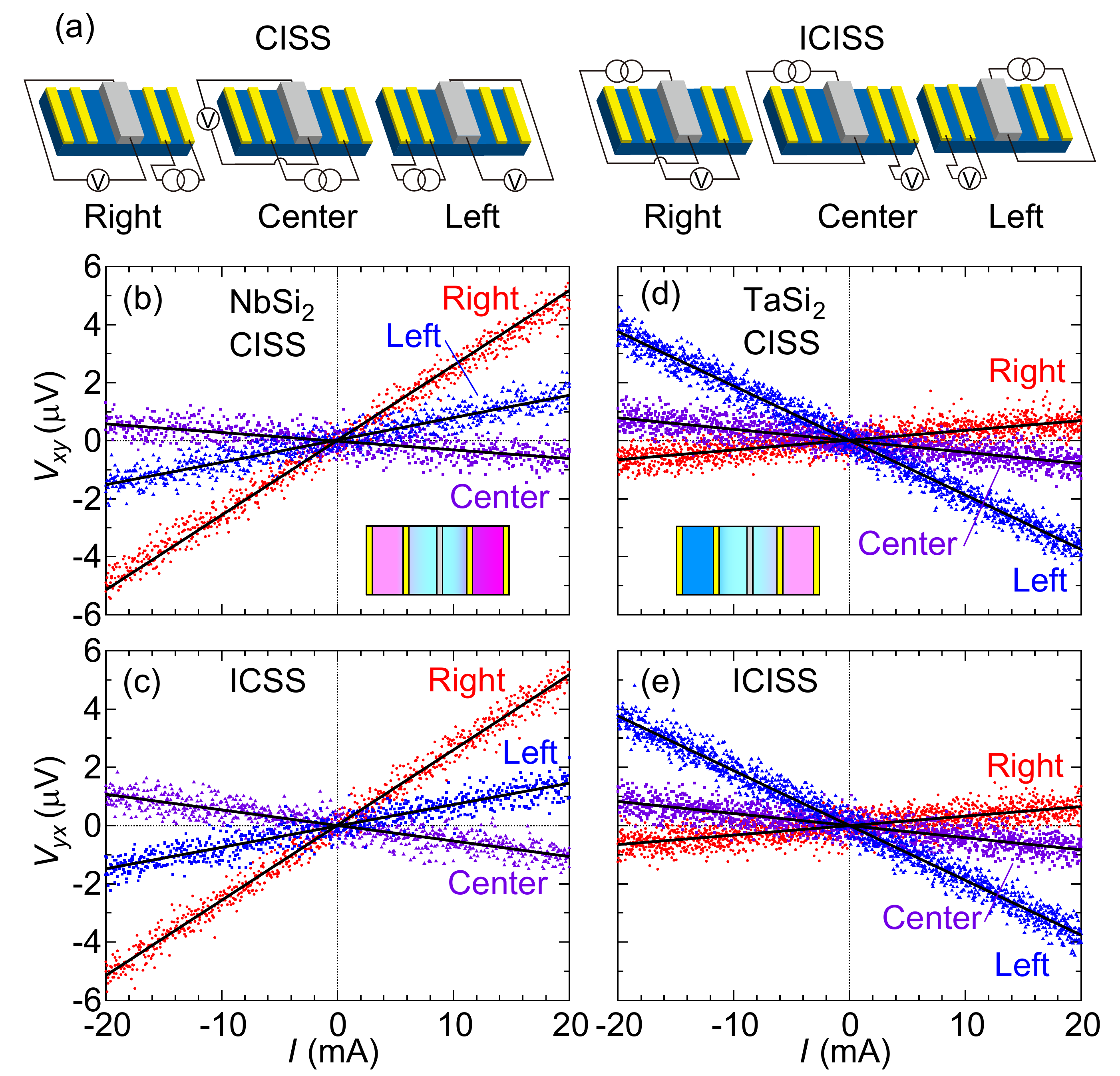}
\caption{Location-sensitive measurements of the direct and inverse CISS in $\mathrm{NbSi_{2}}$ and $\mathrm{TaSi_{2}}$ polycrystals. The measurement setups are shown in (a), while the $I$--$V$ characteristics are presented in (b) to (e). 
All measurements were performed at room temperature without magnetic fields.
The insets in (b) and (d) show schematic drawings of the distribution of chirality grains in the samples.
}
\label{fig3}
\end{figure}

%%%%%%%%%%%%%%%%%%%%%%%%%%%%%%%%%%%%%%%%%%%%%%%%%%%%%%%%%%%%%%%%%%%%%%%%

$\mathrm{NbSi_{2}}$ and $\mathrm{TaSi_{2}}$ compounds crystalize in a hexagonal C20 structure with the space group of $P6_{2}22$ or $P6_{4}22$. 
The handedness of the $\mathrm{NbSi_{2}}$ and $\mathrm{TaSi_{2}}$ crystals can be determined macroscopically by means of precise X-ray diffraction measurements. 
According to the literature~\cite{Sak05}, the former prefers to form a right-handed crystalline structure which belongs to $P6_{2}22$, while the latter in favor of the opposite one in $P6_{4}22$, as illustrated in Fig.~\ref{fig1}(a).

\begin{table*}
\caption{
A list of the conversion coefficient of the direct and inverse CISS signals as well as the longitudinal resistance for the polycrystalline samples of NbSi$_2$ and TaSi$_2$. All the data was obtained at room temperature without magnetic fields.
The slope values for the longitudinal voltage $V_{xx}$, direct CISS voltage $V_{\rm xy}$, and inverse CISS voltage $V_{yx}$ are provided with regard to the applied current $I$. The data shown in Figs.~2 and~3 was obtained with the \#1 samples of NbSi$_2$ and TaSi$_2$.\\
}
\scalebox{1}{ 
\begin{tabular}{ccccccccccc}
\hline
\hline
\multirow{2}{*}{Crystal} & & & \multirow{2}{*}{$V_{xx}/I$\,(\si{\milli\ohm})} & & & \multicolumn{5}{c}{$V_{xy}/I$\,(\si{\milli\ohm})}  \\
\cline{6-11}
 & & & %Center 
  & & & Whole & Left & Center & Right & Sum \\
\hline
\multirow{3}{*}{NbSi$_2$} &\#1 & & 0.9357(10) & & & \,\,0.2879(10) & \,\,0.0771(10) & -0.0299(10) & \,\,0.2581(10) & \,\,0.3053(17) \\
 &\#2 & & 0.46411(8) & & & \,\,0.01050(7) & -0.12695(6) & \,\,0.11455(7) & \,\,0.02375(6) & \,\,0.01135(11) \\ 
 &\#3 & & 0.61627(6) & & & \,\,0.25663(5) & \,\,0.16117(5) & \,\,0.08455(8) & \,\,0.02092(5) & \,\,0.26664(11) \\ 
%\hline
\multirow{3}{*}{TaSi$_2$} &\#1 & & 1.0803(7) & & & -0.1876(7) & -0.1878(7) & -0.0396(7) & \,\,0.0338(7) & -0.1936(12) \\ 
 &\#2 & & 0.6691(7) & & & -0.1207(7) & \,\,0.1323(4) & -0.2496(7) & -0.0067(4) & -0.1250(9) \\ 
 &\#3 & & 0.9903(7) & & & -0.1275(7) & \,\,0.0695(7) & -0.1932(7) & -0.0100(7) & -0.1337(12) \\ 
\hline
\hline
 & & & & & & \multicolumn{5}{c}{$V_{yx}/I$\,(\si{\milli\ohm})} \\
 \cline{6-11}
 & & & & & & Whole & Left & Center & Right & Sum \\
 \hline
\multirow{3}{*}{NbSi$_2$} &\#1 & & & & & \,\,0.2802(10) & \,\,0.0732(10) & -0.0535(10) & \,\,0.2581(10) & \,\,0.2778(17) \\
&\#2 & & & & & \,\,0.01057(6) & -0.12704(7) & \,\,0.11383(7) & \,\,0.02416(6) & \,\,0.01095(12) \\
&\#3 & & & & & \,\,0.25781(5) & \,\,0.16038(6) & \,\,0.08388(5) & \,\,0.01622(5) & \,\,0.26048(9) \\
%\hline
\multirow{3}{*}{TaSi$_2$} &\#1 & & & & & -0.1920(7) & -0.1883(7) & -0.0415(7) & \,\,0.0325(7) & -0.1973(12) \\
&\#2 & & & & & -0.1239(4) & \,\,0.1316(4) & -0.2574(4) & -0.0077(4) & -0.1335(7) \\
&\#3 & & & & & -0.1284(7) & \,\,0.0691(7) & -0.1953(7) & -0.0097(7) & -0.1359(12) \\
\hline
\hline
 
\end{tabular}
}
\label{table1}
\end{table*}
%%%%%%%%%%%%%%%%%%%%%%%%%%%%%%%%%%%%%%%%%%%%%%%%%%%%%%%%%%%%%%%%%%%%

For obtaining polycrystalline ingots, compound powders with a purity of 2N for $\mathrm{NbSi_{2}}$ and 3N for $\mathrm{TaSi_{2}}$, provided by Kojundo Chemical Lab. Co. Ltd., were arc melted in Ar atmosphere by an arc furnace. 
X-ray diffraction patterns of the bulk $\mathrm{NbSi_{2}}$ and $\mathrm{TaSi_{2}}$ are shown in Fig.~\ref{fig1}(b).
Many peaks were observed in the diffraction patterns, as seen for powder samples.
Finite contamination of Nb$_5$Si$_3$ and Ta$_5$Si$_3$ were found as indicated by asterisks. 
It was difficult to find any specific orientation of the planes along which the samples predominantly grow throughout surfaces. 
The mean size of the grains $D$ was calculated as 24 and \SI{27}{\nano \metre} for NbSi$ _2$ and TaSi$_2$, respectively, by applying a Debye-Scherrer formula $D = 0.89\lambda/\beta\cos\theta$ for the 111 peaks. Here, $\lambda$ is the X-ray wavelength and $\beta$ is the full width at half-maximum. 
All the features observed in the diffraction patterns support polycrystallinity of the samples.

Three pieces of polycrystals were cut from the bulk ingots of $\mathrm{NbSi_{2}}$ and $\mathrm{TaSi_{2}}$ for the CISS measurements.
The samples have a rectangular shape with typical dimensions of several millimeters in length and width and one millimeter in depth, as shown in Fig.~\ref{fig1}(c). 
For the detection of spin polarization, platinum (Pt) was deposited by using a magnetron sputter (MSP-1S, Vacuum Device Inc.). A platinum electrode of \SI{0.5}{\milli\metre} in width and \SI{50}{\nano\metre} in thickness was formed on the center of the sample through a metal mask. 
For the charge current injection, gold wires were attached to the sample via a silver paste.

All the electrical transport measurements were performed without magnetic fields. The procedure of the CISS measurements is schematically drawn in Fig.~\ref{fig1}(d).
With an assumption that the CISS effect occurs in a chiral polycrystalline sample, the measurements proceed as follows: 1)~A longitudinal electrical current $I_{xx}$ is applied to the chiral samples and induces the spin polarization because of the CISS response. 2)~Spin-polarized electrons are absorbed into the Pt electrode because of the difference of spin-dependent chemical potential between the chiral sample and Pt electrode. 3)~In the Pt electrode, the spin current is converted to the transverse electrical current $I_{xy}$ via an inverse spin Hall effect~\cite{Val06, Sai06, Kim07}. 
Finally, $I_{xy}$ generates the transverse voltage $V_{xy}$ via the Pt resistance. Therefore, the CISS signals can be detected by measuring $V_{xy}$ as a function of $I_{xx}$.

The inverse CISS signals are detectable in the same samples in the following steps. 1) The transverse electrical current is applied to the Pt electrode. 2) The spin current is generated in a direction orthogonal to the charge current because of a spin Hall effect and diffuses into the chiral sample. 3) The spin-polarized charge current flows in the chiral sample via the inverse CISS and is detected as a longitudinal voltage $V_{yx}$.

Figure~\ref{fig2} shows $I$--$V$ characteristics of the $\mathrm{NbSi_{2}}$ and $\mathrm{TaSi_{2}}$ bulk polycrystals at room temperature. The longitudinal voltage $V_{xx}$ follows the Ohm's law in a current range of \SI{\pm20}{\milli\ampere} in both samples, as shown in Figs.~\ref{fig2}(a) and~\ref{fig2}(d).
Electrical resistivity at room temperature is estimated as \SI{75}{\mu\Omega~\centi\metre} for the $\mathrm{NbSi_{2}}$ polycrystal and \SI{140}{\mu\Omega~\centi\metre} for the $\mathrm{TaSi_{2}}$ one. 
These values are slightly larger than those for the single crystals.

Figure~\ref{fig2}(b) shows the CISS voltage $V_{xy}$ as a function of the applied current $I_{xx}$ in the $\mathrm{NbSi_{2}}$ polycrystalline sample. 
The inverse voltage $V_{yx}$ with the charge current applied into the Pt electrode is presented in Fig.~\ref{fig2}(c).
A linear dependence of the $I$--$V$ characteristics was observed in both measurements. 
Importantly, the polarity of $V_{yx}$ is the same as that of $V_{xy}$ with almost the same amplitude of the signals.

Similar behavior was observed in the $V_{xy}$ and $V_{yx}$ signals in the $\mathrm{TaSi_{2}}$ polycrystalline sample, as shown in Figs.~\ref{fig2}(e) and~\ref{fig2}(f), respectively. 
The other four samples of $\mathrm{NbSi_{2}}$ and $\mathrm{TaSi_{2}}$ prepared for the CISS measurements showed reproducible results.
The present results obtained in the polycrystals are consistent with those for the single crystals~\cite{Shi21}.  
Thus, the data in Fig.~\ref{fig2} indicates that the $\mathrm{NbSi_{2}}$ and $\mathrm{TaSi_{2}}$ polycrystals exhibit the direct and inverse CISS effects and satisfy a reciprocal relationship between them.

To obtain further evidence for the CISS response in bulk polycrystals, location-sensitive CISS measurements were performed, as illustrated in Fig.~\ref{fig3}(a). 
In the nonlocal setup, the intrinsic transverse signal is distinguishable from the extrinsic one because no longitudinal signal due to a misalignment of electrodes is supposed to be superimposed to the signal. 
This situation contrasts with the `whole' measurements shown in Fig.~\ref{fig2}, in which the extrinsic longitudinal signals are hardly excluded. 
In addition, the location-sensitive CISS measurements are good at detecting a distribution of chirality in the sample. 
Furthermore, the location-sensitive CISS measurements may reveal another interesting feature of the CISS. In the case of $\mathrm{NbSi_{2}}$ and $\mathrm{TaSi_{2}}$ single crystals~\cite{Shi21}, it was demonstrated that 
the conversion coefficient of the direct CISS effect $V_{xy}/I$ for the `whole' measurements became almost the same with the summation of those for `left', `center', and `right' measurements, as schematically drawn in Fig.~\ref{fig3}(a). 
This sum rule would be a good indicator of whether a long-range CISS response occurs in the sample.

Figure~\ref{fig3} shows nonlocal and local signals of $V_{xy}$ for the direct CISS as well as $V_{yx}$ signals for the inverse CISS in the $\mathrm{NbSi_{2}}$ and $\mathrm{TaSi_{2}}$ polycrystalline samples at room temperature. Note that the nonlocal measurements were performed in the setup of electrodes with a separation of millimeters in length between the current and voltage contacts.

Now, it is clear that nonlocal signals of $V_{xy}$ and $V_{yx}$ appear in the $\mathrm{NbSi_{2}}$ and $\mathrm{TaSi_{2}}$ polycrystals. 
The slope $V_{xy}/I$ changes, depending on the locations where the charge current is applied, 
as labelled as `left', `center' and `right' in Fig.~\ref{fig3}(a).
The slope values are summarized in Table~\ref{table1}. 

Interestingly, the sum of $V_{xy}/I$ over the three regions in the above measurements almost coincides with the $V_{xy}/I$ value obtained for the `whole' measurement in Fig.~\ref{fig2}. 
Slight deviations are found between them, which may be caused by some additional signals in `whole' and `center' measurements due to the misalignment of the Pt electrode.  

Similar behavior is found with regard to $V_{yx}/I$ in the location-sensitive inverse CISS  measurements. 
A reproducibility of the experimental results including the sum rule was confirmed in the other four samples of NbSi$_2$ and TaSi$_2$ polycrystals, as summarized in Table~\ref{table1}.

%%%%%%%%%%%%%%%%%%%%%%%%%%%%%%%%%%%%%%%%%%%%%%%%%%%%%%%%%%%%%%%%%%%%%%%%
\begin{figure}[t]
\includegraphics[bb= 25 0 582 756, width=0.8\linewidth]{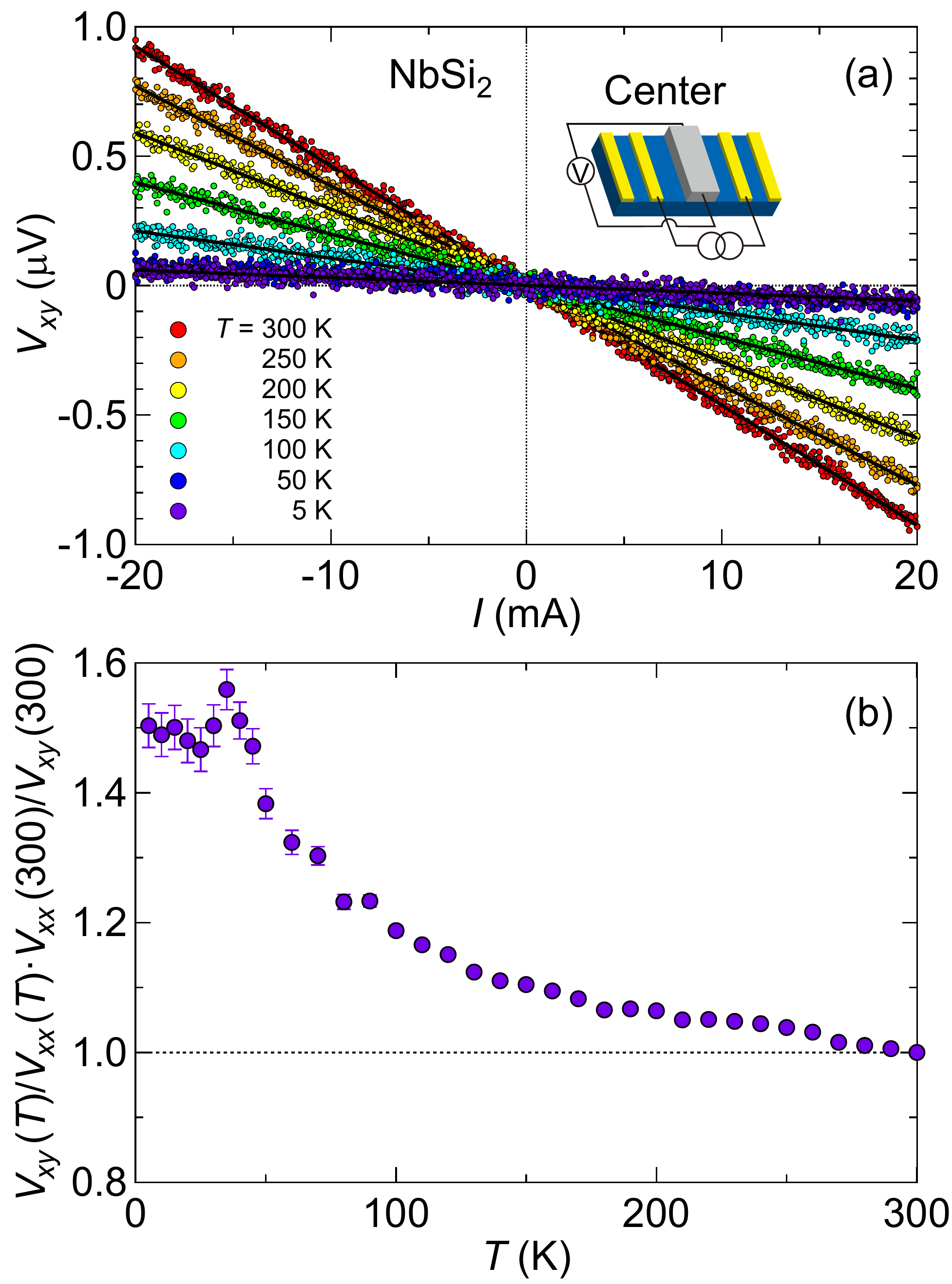}
\caption{Temperature variation of the direct CISS in $\mathrm{NbSi_{2}}$ with `center' location. (a) $I$--$V_{xy}$ characteristics at various temperatures. (b) Temperature dependence of the signal ratio of $V_{xy}$ and $V_{xx}$ normalized at 300\,K. 
}
\label{fig4}
\end{figure}

%%%%%%%%%%%%%%%%%%%%%%%%%%%%%%%%%%%%%%%%%%%%%%%%%%%%%%%%%%%%%%%%%%%%%%%%

We demonstrate that the $V_{xy}$ signals result from the CISS effect rather than a misalignment of the electrodes, by comparing the temperature dependence of $V_{xy}$ with that of $V_{xx}$.
The direct CISS data in the `center' measurement is adopted, where the current is applied in the same region as the $V_{xx}$ measurement, so as to exclude a possible difference in the locational dependence of the voltage signals as a function of temperature.

Figure~\ref{fig4}(a) shows the $I$--$V_{xy}$ curves in the `center' region of NbSi$_2$ at various temperatures.   
A linear dependence remains down to the lowest measured temperature of 5\,K,  and the conversion coefficient reduces with decreasing temperature.  
In the CISS scenario, $V_{xy}$ is a product of the combined resistance of metallic Pt and NbSi$_2$ and the charge current, generated by the spin current via the inverse spin Hall effect. 
The former will reduce at low temperatures, while the latter will increase. 
Thus, the observed reduction of $V_{xy}$ is caused mainly by the resistance. 
In this connection, we regard that the temperature variation of the ratio of transverse $V_{xy}$ and longitudinal $V_{xx}$ signals will be useful for clarifying the origin, as shown in the following.  

The temperature dependence of the signal ratio of $V_{xy}$/$V_{xx}$ normalized at 300\,K  is shown in Fig.~\ref{fig4}(b).   
Measurements were performed after stabilizing temperature with an accuracy of 0.02\,K to prevent the errors caused by the temperature difference between the CISS and longitudinal voltage measurements.    
The normalized $V_{xy}$/$V_{xx}$ gradually increases with decreasing temperature, and shows a maximum at around 35\,K, and takes the constant value below 20\,K.

If $V_{xy}$ originated from the misalignment, the normalized $V_{xy}$/$V_{xx}$ should be the constant in the whole temperature range, as denoted by the dashed line.    
However, a significant difference appears in the temperature dependence between $V_{xy}$ and $V_{xx}$, indicating that the $V_{xy}$ signal is not caused by the misalignment of electrodes but by the CISS effect.

Regarding a locational dependence of the signal polarity, it should reflect a distribution of the handedness in polycrystalline samples. 
Indeed, it was found that crystalline grains with different chirality exist even in the crystals that were supposed to be of single handedness on the basis of macroscopic analysis using x-ray~\cite{Sak05}. 
To clarify a correlation between the signal polarity and handedness in polycrystals needs further investigations.

We demonstrate the direct and inverse CISS responses in millimeter-sized polycrystalline samples of chiral disilicide $\mathrm{NbSi_{2}}$ and $\mathrm{TaSi_{2}}$. 
The transverse responses exhibit a temperature dependence, which is  different from that of the longitudinal signals and regarded as an indication of the CISS origin. 
Handedness of the samples was determined on average even in the polycrystals via the electrical transport measurements. 
The emergence of the nonlocal signals in chiral polycrystals requires a very robust protection and propagation of the spin polarization throughout the samples. 
The origin of the spin polarization should be associated with a spin splitting band structure, which was already observed in chiral disilicide crystals~\cite{Onu14}. 
However, the mechanism for generating robust CISS signals over millimeters remains to be clarified.

On the basis of the experimental results obtained in the present study, we propose that chiral polycrystals can work as a good conductor of spins with a very tiny loss of the spin polarization. 
Indeed, chiral polycrystals could transport spins over long distances irrespective of their crystallinity. 
The materials examined in this study contains 4$d$- or 5$d$-transition metals, which have strong spin-orbit coupling~\cite{Onu14}. However, it turns out that even such materials can exhibit a long-range transportation of the spin polarizations in spite of relatively short spin diffusion lengths of the materials. Thus, a high degree of freedom in material choice may be guaranteed for chiral materials utilized as a spin conductor.

See the supplementary material for the reference experiment using an achiral material.

\begin{acknowledgments}
We thank Jun-ichiro Ohe, Hiroshi M. Yamamoto, Jun-ichio Kishine, Yusuke Kato, and Takuya Sato for useful discussions. We acknowledge support from Grants-in-Aid for Scientific Research (No. 17H02767, No. 17H02923, No. 19K03751 and No. 	21H01032) and Research Grant of Specially Promoted Research Program by Toyota RIKEN. 

\end{acknowledgments}

\section*{Conflict of Interest}
The authors have no conflicts to disclose.

\section*{Data Availability Statement}
The data that support the findings of this study are available from the corresponding author upon reasonable request.

\nocite{*}
%\bibliography{aipsamp}% Produces the bibliography via BibTeX.

\begin{thebibliography}{99}\label{sec:TeXbooks}%

% Papers at the beginning
\bibitem{Goh11} 
B. G\"{o}hler, V. Hamelbeck, T. Z. Markus, M. Kettner, G. F. Hanne, Z. Vager, R. Naaman, H. Zacharias, Science \textbf{331}, 894 (2011).
\bibitem{Xie11}
Z. Xie, T. Z. Markus, S. R. Cohen, Z. Vager, R. Gutierrez, R. Naaman, Nano Lett. \textbf{11}, 4652 (2011).

% Review
\bibitem{Naa12} 
R. Naaman, and D. H. Waldeck, J. Phys. Chem. Lett. \textbf{3}, 21782187 (2012).
\bibitem{Naa15}
R. Naaman, D. H.Waldeck, Annu. Rev. Phys. Chem. \textbf{66}, 263281 (2015).

\bibitem{Wal21}
D. H. Waldeck, R. Naaman, and Y. Paltiel, APL Mater. \textbf{9}, 040902 (2021).

%Suplermolecule
\bibitem{Ros13}
R. A. Rosenberg, J. M. Symonds, V. Kalyanaraman, T. Markus, T. M. Orlando, R. Naaman, E. A. Medina, F. A. L\'{o}pez, V. Mujica, J. Phys. Chem. C \textbf{117}, 22307 (2013).
\bibitem{Kum13}
K. S. Kumar, N. Kantor-Uriel, S. P. Mathew, R. Guliamov, R. A. Naaman, Phys. Chem. Chem. Phys. \textbf{15}, 18357 (2013).

%peptide
\bibitem{Dor13} 
O. B. Dor, S. Yochelis, S. P. Mathew, R. Naaman, Y. Paltiel, Nat. Commun. \textbf{4}, 2256 (2013).
%\bibitem{Dor14}
%O. B. Dor, N. Morali, S. Yochelis, L. T. Baczewski, Y. Paltiel, Nano Lett. \textbf{14}, 6042 (2014).
\bibitem{Ket15}
M. Kettner, B. G\"{o}hler, H. Zacharias, D. Mishra, V. Kiran, R. Naaman, C. Fontanesi, D. H. Waldeck, S. S\k{e}k, J. Paw{\l}owski, J. Juhaniewicz, J. Phys. Chem. C \textbf{119}, 14542 (2015).

%QDs
\bibitem{Ara17}
A. C. Aragon\`{e}s, E. Medina, M. Ferrer-Huerta, N. Gimeno, M. Teixid\'{o}, J. L. Palma, N. Tao, J. M. Ugalde, E. Giralt, I. D\'{i}ez-P\'{e}rez, and V. Mujica, Small \textbf{13}, 1602519 (2017).
\bibitem{Kop17}
G. Koplovitz, D. Primc, D. O. Ben, S. Yochelis, D. Rotem, D. Porath, Y. Paltiel, Adv. Mater. \textbf{29}, 1606748 (2017).
\bibitem{Tas18}
F. Tassinari, D. R. Jayarathna, N. Kantor-Uriel, K. L. Davis, V. Varade, C. Achim, R. Naaman, Adv. Mater. \textbf{30}, 1706423 (2018).
\bibitem{Var18}
V. Varade, T. Markus, K. Vankayala, N. Friedman, M. Sheves, D. H. Waldeck, R. Naaman, Phys. Chem. Chem. Phys. \textbf{20}, 1091 (2018).

%Amino acid
\bibitem{Zha19}
W. Zhang, J. Li, G. Lu, H. Guana, L. Haoa, Chem. Commun. \textbf{55}, 13390 (2019). 
\bibitem{Ziv19} 
A. Ziv, A. Saha, H. Alpern, N. Sukenik, L. T. Baczewski, S. Yochelis, M. Reches, Y. Paltiel, Adv. Mater. \textbf{31}, 1904206 (2019).
\bibitem{Mis19}
S. Mishra, S. Pirbadian, A. K. Mondal, M. Y. El-Naggar, R. Naaman, J. Am. Chem. Soc. \textbf{141}, 19198 (2019).

%Helicene
\bibitem{Kir16}
V. Kiran, S. P. Mathew, S. R. Cohen, I. H. Delgado, J. Lacour, and R. Naaman, Adv. Mater. \textbf{28}, 1957, (2016). 
\bibitem{Ket18}
M. Kettner, V. V. Maslyuk, D. Nurenberg, J. Seibel, R. Gutierrez, G. Cuniberti, K. Ernst, H. Zacharias, J. Phys. Chem. Lett. \textbf{9}, 2025 (2018).

%Molecule moter
\bibitem{Sud19}
M. Suda, Y. Thathong, V. Promarak, H. Kojima, M. Nakamura, T. Shiraogawa, M. Ehara, H. M. Yamamoto, Nat. Commun. \textbf{10}, 2455 (2019).
\bibitem{Kul20}
C. Kulkarni, A. K. Mondal, T. K. Das, G. Grinbom, F. Tassinari, M. F. J. Mabesoone, E. W. Meijer, R. Naaman, Adv. Mater. \textbf{32}, 1904965 (2020). 



\bibitem{Inu20}
A. Inui, R. Aoki, Y. Nishiue, K. Shiota, Y. Kousaka, H. Shishido, D. Hirobe, M. Suda, J. Ohe, J. Kishine, H. Yamamoto, Y. Togawa, Phys. Rev. Lett. \textbf{124}, 16602 (2020).
\bibitem{Nab20}
Y. Nabei, D. Hirobe, Y. Shimamoto, K. Shiota, A. Inui, Y. Kousaka, Y. Togawa, and H. M. Yamamoto, Appl. Phys. Lett. \textbf{117}, 052408 (2020).
%Current-induced bulk magnetization of a chiral crystal CrNb3S6, 
\bibitem{Shi21}
K. Shiota, A. Inui, Y. Hosaka, R. Amano, Y. \={O}nuki, M. Hedo, T. Nakama, D. Hirobe, J. Ohe, J. Kishine, H. M. Yamamoto, H. Shishido, and Y. Togawa, Phys. Rev. Lett. \textbf{127}, 126602 (2021).

\bibitem{Sak05}
H. Sakamoto, A. Fujii, K. Tanaka, H. Inui, Acta Mater. \textbf{53}, 41 (2005).

\bibitem{Val06}
S. O. Valenzuela and M. Tinkham, Naure \textbf{442}, 176 (2006).
\bibitem{Sai06}
E. Saitoh, M. Ueda, H.Miyajima, G.Tatara, Appl. Phys. Lett. \textbf{88}, 182509 (2006).
\bibitem{Kim07}
T. Kimura, Y. Otani, T. Sato, S. Takahashi, S. Maekawa, Phys. Rev. Lett. \textbf{98}, 156601 (2007).


\bibitem{Onu14}
Y. \={O}nuki, A. Nakamura, T. Uejo, A. Teruya, M. Hedo, T. Nakama, F. Honda, H. Harima J. Phys. Soc. Jpn. \textbf{83}, 061018 (2014).


\end{thebibliography}

%%%%%%%%%%%%%%%%%%%%%%%%%%%%%%%%%%%%%%%%%%%%%%%%%%%%%%%%%%%%%%%%%%%%%%%%%%%%%%%%%%%%%%%%%%%%%%%%%%%%%%%%%%%%%%%%%%%%%%%%%%%%%%%%%%%%%%%%%%%%%%%%%%%%%%%%
%%%%%%%%%%%%%%%%%%%%%%%%%%%%%%%%%%%%%%%%%%%%%%%%%%%%%%%%%%%%%%%%%%%%%%%%%%%%%%%%%%%%%%%%%%%%%%%%%%%%%%%%%%%%%%%%%%%%%%%%%%%%%%%%%%%%%%%%%%%%%%%%%%%%%%%%
\newpage
\begin{widetext}
\section{Supplementary materials : Detection of chirality-induced spin polarization over millimeters in polycrystalline bulk samples of 
chiral disilicides $\mathrm{NbSi_{2}}$ and $\mathrm{TaSi_{2}}$}

\setcounter{section}{0}
\setcounter{equation}{0}
\setcounter{figure}{0}
\setcounter{table}{0}

\renewcommand{\thesection}{S\arabic{section}}
\renewcommand{\theequation}{S\arabic{equation}}
\renewcommand{\thefigure}{S\arabic{figure}}
\renewcommand{\thetable}{S\arabic{table}}
\renewcommand{\bibnumfmt}[1]{[S#1]}
\renewcommand{\citenumfont}[1]{S#1}
\makeatletter

\section{Reference experiment using an achiral material}

Regarding the reference experiment using an achiral material, electrical properties were examined in Tantalium (Ta) samples made of a polycrystalline foil. A Ta element exhibits a W-type crystalline structure with the space group of $Im$\={3}$m$. We do not expect the spin polarization in Ta samples because of an achiral nature of the material. Thus, the detected signals should be caused by extrinsic effects such as a misalignment of electrodes.

%%%%%%%%%%%%%%%%%%%%%%%%%%%%%%%%%%%%%%%%%%%%%%%%%%%%%%%%%%%%%%%%%%%%
\begin{figure}[h]
\includegraphics[width=0.8\linewidth]{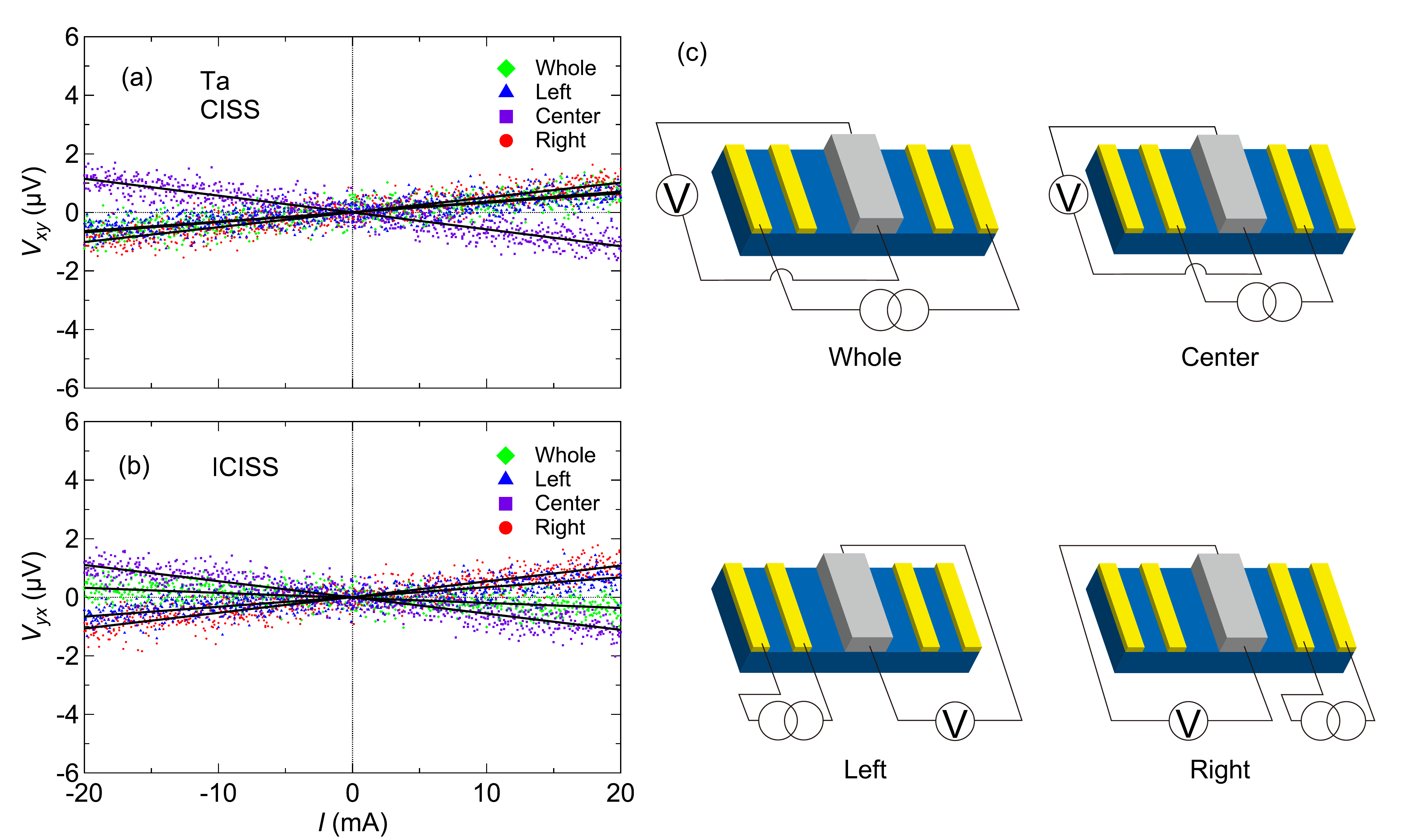}
\caption{Location-sensitive measurements of $V_{xy}$ (a) and $V_{yx}$ (b) for the polycrystalline Ta sample. The measurement setups are drawn in (c). Solid lines indicate linear regression fits. }
\label{figS1}
\end{figure}
%%%%%%%%%%%%%%%%%%%%%%%%%%%%%%%%%%%%%%%%%%%%%%%%%%%%%%%%%%%%%%%%%%%%

The rectangular samples with a dimension of \SI{5}{\milli \metre} in length and \SI{4}{\milli \metre} in width were cut  from a Ta foil of \SI{0.1}{\milli \metre} in thickness with a purity of 99.95\% for electrical measurements.
X-ray experiments revealed that the foils used in this study consist of many crystalline grains. The samples have the same configuration of electrodes as that used for the CISS and ICISS measurements in the polycrystalline samples of chiral NbSi$_2$ and TaSi$_2$, presented in the main manuscript. 

Figure~\ref{figS1} shows current and voltage characteristics obtained in the Ta polycrystaline sample.
The measurement setups are schematically drawn in Fig.~\ref{figS1}(c), which are the same used for the NbSi$_2$ and TaSi$_2$ samples, as presented in Fig.~2(a) in the main text. While the $V_{xy}$ and $V_{yx}$ signals are small and scattered, the slope $V_{xy}/I$ and $V_{yx}/I$ are evaluated by using linear regression fits. The obtained values are summarized in Table~\ref{table2}. 

The maximum amplitude of the slope of the $V_{xy}$ and $V_{yx}$ signals for the Ta sample is one order of magnitude smaller than those observed in the chiral disilicide samples. 
As described above, the transverse signal should be originated from a misalignment of the detection electrode. 
In this case, the signal is proportional to the longitudinal resistance and thus inversely proportional to the sample thickness. 
Note that the thickness of Ta samples is one order of magnitude thinner than that for the chiral disilicide samples presented in the main manuscript. 
Naively speaking, extrinsic effects for the transverse voltages for the chiral disilicide samples would be much smaller than
those for the Ta sample. 
Therefore, we can safely conclude that detected signals of $V_{xy}$ and $V_{yx}$ in the NbSi$_2$ and TaSi$_2$ samples are due to the intrinsic CISS effect.

The sum rule is unlikely to hold in the Ta samples, as shown in Table~\ref{table2}. This behavior is in sharp contrast with the experimental finding that the sum rule is well satisfied in the chiral disilicide samples, as described in the main text. We regard the observations in the NbSi$_2$ and TaSi$_2$ samples as a sign of the CISS effect that chiral materials exhibit.

%%%%%%%%%%%%%%%%%%%%%%%%%%%%%%%%%%%%%%%%%%%%%%%%%%%%%%%%%%%%%%%%%%%%
\begin{table*}
\caption{The slope values of $V_{xy}/I$ and $V_{yx}/I$ for the Ta sample are summarized. The sequence of the data is along with that presented in the main text for NbSi$_2$ and TaSi$_2$.}
\scalebox{1}{ 
\begin{tabular}{ccccccccccc}
\hline
\hline
\multirow{2}{*}{Crystal} & & \multirow{2}{*}{$V_{xx}/I$\,(\si{\milli\ohm})} & & & \multicolumn{5}{c}{$V_{xy}/I$\,(\si{\milli\ohm})} \\
\cline{6-11}
 & & & & & & Whole & Left & Center & Right & Sum \\
\hline
Ta & & 2.4582(9) & & & 0.0349(9) & 0.0323(9) & -0.0575(9) & 0.0507(9) & 0.0255(16) \\
\hline
\hline
 & & & & & & \multicolumn{5}{c}{$V_{yx}/I$\,(\si{\milli\ohm})} \\
\cline{6-11}
 & & & & & & Whole & Left & Center & Right & Sum \\
\hline
Ta & & & & & -0.0172(9) & 0.0335(9) & -0.0554(10) & 0.0536(10) & 0.0317(17) \\
\hline
\hline
\end{tabular}
}
\label{table2}
\end{table*}
%%%%%%%%%%%%%%%%%%%%%%%%%%%%%%%%%%%%%%%%%%%%%%%%%%%%%%%%%%%%%%%%%%%%

\end{widetext}

\end{document}